\documentclass[twocolumn,showpacs,showkeys,prbamsmath,amssymb]{revtex4}
\usepackage{graphicx}% Include figure files
\usepackage{dcolumn}% Align table columns on decimal point
\usepackage{bm}% bold math

\begin{document}

\title{Phase Diagram of mixed bond Ising systems by use of Monte Carlo and the effective-field theory}
\author{J. B. Santos-Filho$^{1}$, N. O. Moreno$^{2}$, Douglas F. de Albuquerque$^{3}$}
\affiliation{$^{1}$Departamento de F\'{\i}sica, Universidade Federal
de Minas Gerais, 31270-901, Belo Horizonte, MG, Brasil}
\affiliation{$^{2}$Departamento de F\'{\i}sica, Universidade Federal
de Sergipe, 49100-000, S\~ao Cristov\~ao, SE, Brasil}
\affiliation{$^{3}$Departamento de Matem\'atica, Universidade
Federal de Sergipe, 49100-000, S\~ao Cristov\~ao, SE, Brasil}
\date{\today}

\begin{abstract}
The phase transition of a random mixed-bond Ising ferromagnet 
on a cubic lattice model is studied both numerically and analytically. 
In this work, we use the Cluster algorithms of Wolff
and Glauber to simulate the dynamics of the system. We obtained the 
thermodynamic quantities such as magnetization, susceptibility, and specific heat. 
Our results were compared with those obtained using a new technique in effective field theory
that employs similar probability distribution within the framework
of two-site clusters.

\end{abstract}



\maketitle

\section{INTRODUCTION}

The study of the effects of disorder in magnetic systems has been
an object of intense investigations during the last five decades.

The Monte Carlo technique is an useful tool which in many cases, 
gives better results regarding other methods from analytical approximations. 
The influence of quenched, random disorder on phase transitions is of great
importance in a large variety of fields \cite{Folk}. For pure
systems exhibiting a continuous phase transition, Harris
\cite{Harris} derived the criterion that random disorder is a
relevant perturbation when the exponent of the specific heat of
the pure system is positive, $\alpha > 0$. In this case one
expects that the system falls into a new universality class
with critical exponents governed by a disordered fixed point.
For $\alpha < 0$ disorder is irrelevant, and in the marginal
case $\alpha = 0$ no prediction can be made.

Since for the three-dimensional (3D) Ising model it is
well known that $\alpha > 0$, quenched, random disorder should be
relevant for this model. In three dimensions (3D) most of the computer
simulation studies have concentrated mainly on the site-diluted Ising
model \cite{Wiseman,Ballesteros}.

In this work, we study the Ising model with mixed-bond by using
of Monte Carlo simulation, applying the algorithm cluster of
Wolff \cite{Wolff}.

\section{Model and Simulation Setup}

We study the spin 1/2 ferromagnetic Ising mixed-bond model defined by
the following Hamiltonian

\begin{equation}
\beta \mathcal{H} =  \sum_{\langle {ij} \rangle}K_{ij} \sigma _i
\sigma _j  \ \  \ \  \ \ (\sigma _i= \pm 1),
\end{equation}

\noindent where the sum extends over all pairs of neighboring
sites on a cubic lattice of linear size L with periodic boundary
conditions, $\beta=1/k_B T$ and the exchange couplings $K_{ij}$ are allowed to take two
different values $K_{ij} = K \equiv J/k_{B}T$ and 0. The interactions are assumed 
to be independent random variables with distribution 

\begin{equation}
 P({K_{ij}) = p\delta ({K_{ij}  - K})+({1 - p})\delta ( {K_{ij}  - \lambda
 K}),}
\end{equation}

\noindent where $p$ is the concentration of magnetic bonds
in the system bonds such that $p = 1$ corresponds to the pure case
and $\lambda $ is the competition parameter with $ \left| \lambda  \right| \le 1$.

The simulations were performed on a set of following lattice sizes 
L = 10, 16, 20, 26, 30, 36, 40 with periodic boundary conditions. The aim of 
the first set of simulations is to estimate the critical temperature of
the model at different L. Due to the finite-size scaling theory \cite{Barber}, 
the finite system of linear size $L$ will demonstrate an evidence of a critical behavior 
at a certain temperature $T_{C}(L)$ which differs from the critical temperature 
of the infinite system $T_{C}(\infty)$ \cite{Ferrenberg}.

\begin{equation}
T_C (L) = T_C (\infty ) + \alpha L^{ - 1/v}  + ...,
\end{equation}

\noindent where the correction-to-scaling terms have been omitted.

The static thermodynamic quantities of interest include the
average magnetization $M$ and the magnetic susceptibility $\chi $

\begin{equation}
M = \frac{1}{n}\sum\limits_{i = 1}^n {\sigma _i},
\end{equation}

\begin{equation}
\chi  = \frac{1}{{k_ {B} T}}[{\langle {M ^2} \rangle - \langle M
\rangle ^2 }].
\end{equation}

The phase diagram is obtained numerically from the maxima of a
diverging quantity. Here we choose the magnetic susceptibility,
since the stability of the disordered fixed point implies that the
specific heat exponent is negative in the random
system \cite{Chayes,Zamora}. Thus, the error in this quantity is
larger than for the susceptibility. To get an accurate
determination of the maxima of the susceptibility, we used the
histogram reweighting technique with 2500 Monte Carlo sweeps (MCS)
and between 2500 and 5000 samples of disorder. The number of Monte
Carlo sweeps is justified by the increasing behavior of the energy
autocorrelation time, $\tau _{E}$, and we chose for each size at
least 250 independent measurements of the physical quantities
(N$_{MCS} > 250$ $\tau _{E}$). The choice of N$_{MCS}$ is
justified by the increasing behavior of the energy autocorrelation
time $\tau _{E}$ as a function of $p$ and L. At the critical point
of a second-order phase transition one expects a finite-size
scaling (FSS) behavior $\tau _{E}$ $\propto $ L$^{z}$, where z is
the dynamical critical exponent

\begin{figure}[b]
\includegraphics[scale=0.8]{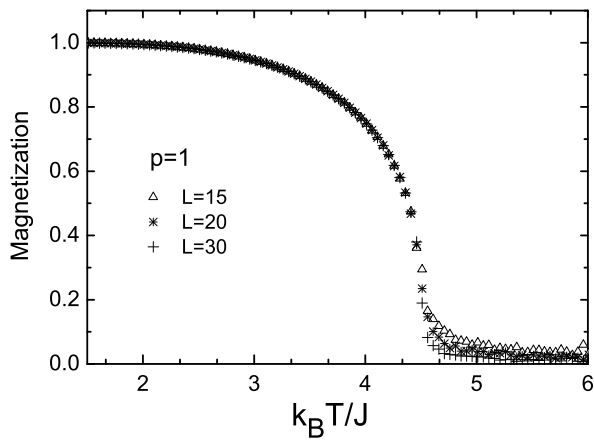}
\caption{\label{magn}Magnetization vs. $k_{B}T/J$  $p=1$ for the 
Ising model with L = 15, 20 and 30 in a cubic lattice.}
\end{figure}

\section{Results}

The figure~\ref{magn} display curves of magnetization versus temperature 
through computational simulation (Wolff) for $p=1$ and $L$ = 15, 20 and 30. 
The critical temperature obtained when $p = 1$ was $T_{C}=4.510$ it
is close of the expected value $T_{C}=4.51$.\cite{Domb} We observe that the curves
keep the same behavior, in spite of in the proximities of the
critical point they move away each other. The critical point was estimated 
of the inflection of the curve. It can be notice that
increasing the lattice size, i.e. the value of L, we get
more precision to estimate the critical temperature.

\begin{figure}[t]
\includegraphics[scale=0.8]{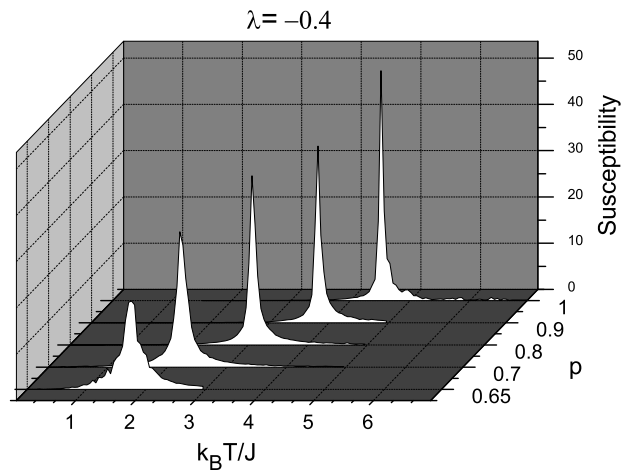}
\caption{\label{susc}$\chi$ vs. $k_{B}T/J$ for the Ising model with
$\lambda = -0.4$ and L = 40 for several concentrations $p$.}
\end{figure}

The magnetic susceptibility as a function of temperature for different $p$ values 
and for $\lambda =-0.4$ and L = 40 is shown in Fig.~\ref{susc}. 
The peaks are sharper for values lower of p. To $p<0.65$ and $\lambda =-0.4$ the susceptibility not displays the
peak associate with the magnetic transition due to competitions of the exchange interactions. 
When $\lambda \geq 0$ the system presents always long range order.
We used the histogram reweighting technique with 2500 Monte Carlo
sweeps (MCS) and between 2500 and 5000 samples of disorder to
get an accurate determination of the maxima of the susceptibility.

\begin{figure}[h]
\includegraphics[scale=0.8]{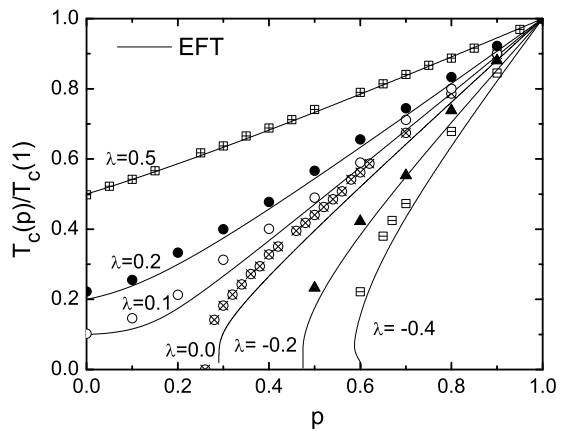}
\caption{\label{fase}Phase diagram of the 3D mixed-bond Ising
model compared with the effective field approximation.}
\end{figure}

The phase diagram obtained from the location of the maxima of the
susceptibility for the largest lattice size (L = 40) as a function
of the concentration of magnetic bonds is shown in Fig.~\ref{fase}
for $\lambda $=0, 0.1, 0.2, 0.5, -0.2 and -0.4. Solid lines are the
predictions of the effective field approximation. A very good agreement
 with the simulated transition line is obtained.

\begin{figure}[t]
\includegraphics[scale=0.8]{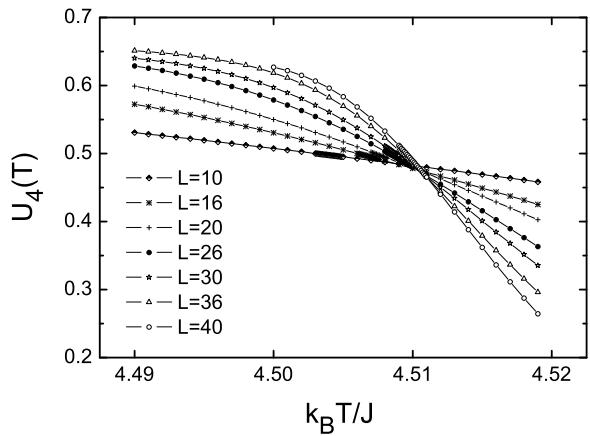}
\caption{\label{cumulante}Fourth-order Binder cumulant $U_{L}$ vs. $k_{B}T/J$, for 
different lattice sizes as indicated in figure.}
\end{figure}

The reduced fourth-order Binder cummulants\cite{Binder} supply an alternative method to estimated
critical points that can be determined from the crossing point of the cumulants for different lattice sizes. 
It is calculated by

\[
U_L  = 1 - \frac{{\left[ {\left\langle {m^4 } \right\rangle } \right]}}{{3\left[ {\left\langle {m^2 } \right\rangle ^2 } \right]}},
\]

\noindent where [...] denotes the average over disorder and $\langle ...\rangle $ refers at the thermal average.
As an example, we show in Fig~\ref{cumulante} the T-dependence of the reduced fourth-order Binder cummulant for $p=1$ and for various lattice sizes. Critical temperature obtained from this figure is in agreement with those obtained of the maxima of the magnetic susceptibility.

\begin{figure}[b]
\includegraphics[scale=0.8]{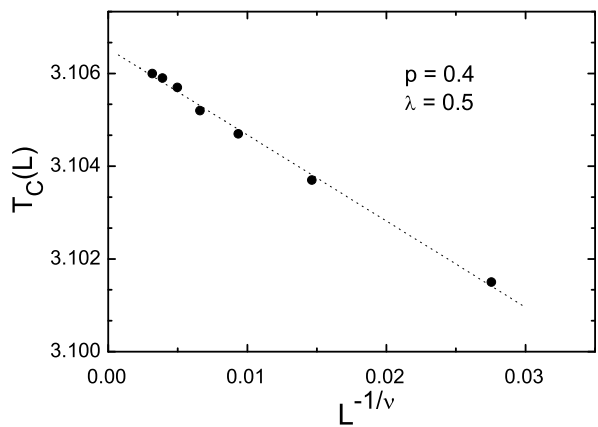}
\caption{\label{tc}Size-dependent critical temperature $T_{C}(L)$vs. $L^{-1/\nu}$, for
$\lambda=0.5$ and $p=0.4$.}
\end{figure}

The peak locations of the maxima susceptibility for each $L$ are plotted versus $L^{-1/\nu}$, where the value of $\nu$
is determined of the linear fit of log-log plot $\frac{{\partial U_L }}{{\partial T}}$ vs $L$ (not shown here). The critical temperature can be estimated from an infinite-size extrapolation in according with Eq.(3). We illustrate this procedure in Fig~\ref{tc} for $\lambda=0.5$ and $p=0.4$, the fit yield to $T_{C} = 3.1065(3)$.

\begin{table}[h]
\caption{\label{table1}Critical exponents for three dilutions and $\lambda=0.5$. }
\begin{ruledtabular}

\begin{tabular}{ccccc}

  % after \\: \hline or \cline{col1-col2} \cline{col3-col4} ...
  p & 1 & 0.8 & 0.6 & 0.4 \\   \hline
  $1/\nu$ & 1.591(7) & 1.68(7) & 1.63(4) & 1.56(3) \\
  $\beta/\nu$ & 0.4920(5) & 0.49(1) & 0.50(13) & 0.496(5) \\
  $\gamma/\nu$ & 2.008(6) & 2.06(3) & 1.950(6) & 2.001(6) \\

\end{tabular}
\end{ruledtabular}
\end{table}

The average magnetization $m$ and the magnetic susceptibility $\chi$ scale with the lattice size as:

\begin{equation}
m\sim a_m L^{ - \beta /\nu }, \  \ \chi \sim a_\chi  L^{\gamma /\nu ,}
\end{equation}

\noindent here $a_m $ and $a_\chi $ are non-universal amplitudes. From these power-laws we extracted the exponents $\beta$ and $\gamma$ plotting in logarithm scale the lattice size dependence of the susceptibility and average magnetization. The critical exponents obtained to p = 0.4, 0.6, 0.8 and $\lambda=0.5$ are shown in Table I. The critical exponents oscillate without having an apparent correlation with the dilution. They are pretty close of those for disordered Ising model \cite{Ballesteros}

\section{Conclusion}

We carried out Monte Carlo simulations for study the influence of bond dilution on the critical
properties of the Ising Model applied for cubic lattice. We
obtained thermodynamic parameters for $ \left| \lambda  \right| \le 1$.

Satisfactory results are obtained using the algorithm of Wolff and showed that this technique 
is appropriated to treat the Mixed-bond problem. The Monte Carlo technique results gives similar results to the obtained ones by the effective field theory.

The critical behavior of the mixed-bond model is governed by the same universality class 
as the site-diluted model and pure Ising model.

\section*{ACKNOWLEDGMENT}

N.O. Moreno is partially supported by FAPITEC-SE and CNPq
Brazilian agencies. J. B. Santos Filho acknowledge financial
support by CAPES.

\end{document}